\documentclass[aps,pra,nofootinbib,twocolumn,floatfix,showpacs,superscriptaddress]{revtex4-1}
\usepackage[utf8]{inputenc}
\usepackage{amsmath,graphicx,epsfig,amsthm,color,bm}
\usepackage{pifont,bbding,amssymb,soul,mathrsfs}
\usepackage{indentfirst}
\usepackage{times,dsfont,units}
\usepackage{appendix, comment}
\usepackage{array}
\usepackage{makecell}
\usepackage{booktabs}
\usepackage{diagbox}
\usepackage{threeparttable}

\usepackage{tikz,pgfplots}

\newcommand{\Tr}{{\rm Tr}}

\newcommand{\bra}[1]{\langle #1|}
\newcommand{\ket}[1]{|#1\rangle}

\newcommand{\comments}[1]{}
\usepackage{lmodern}
\usepackage[T1]{fontenc}
\usepackage[hyperindex,breaklinks,colorlinks=true,citecolor=blue,urlcolor=blue]{hyperref}

\usepackage[capitalize]{cleveref}

\pgfplotsset{compat=1.17} 
\begin{document}
\title{Experimental distillation of tripartite quantum steering with an optimal local filtering operation}
\author{Qian-Xi Zhang}
\affiliation{School of Physics, State Key Laboratory of Crystal Materials, Shandong University, Jinan 250100, China.}

\author{Xiao-Xu Fang}
\affiliation{School of Physics, State Key Laboratory of Crystal Materials, Shandong University, Jinan 250100, China.}

\author{He Lu}
\email{luhe@sdu.edu.cn}
\affiliation{School of Physics, State Key Laboratory of Crystal Materials, Shandong University, Jinan 250100, China.}
\affiliation{Shenzhen Research Institute of Shandong University, Shenzhen 518057, China.}

\begin{abstract}
Multipartite Einstein-Podolsky-Rosen~(EPR) steering admits multipartite entanglement in the presence of uncharacterized verifiers, enabling practical applications in semi-device-independent protocols. Such applications generally require stronger steerability, while the unavoidable noise weakens steerability and consequently degrades the performance of quantum information processing. Here, we propose the local filtering operation that can maximally distill genuine tripartite EPR steering from $N$ copies of three-qubit generalized Greenberger-Horne-Zeilinger states, in the context of two semi-device-independent scenarios. The optimal filtering operation is determined by the maximization of assemblage fidelity. Analytical and numerical results indicate the advantage of proposed filtering operation when $N$ is finite and the steerability of initial assemblages is weak. Experimentally, a proof-of-principle demonstration of two-copy distillation is realized with optical system. The advantage of optimal local filtering operation is confirmed by the distilled assemblage in terms of higher assemblage fidelity with perfectly genuine tripartite steerable assemblages, as well as the greater violation of the inequality to witness genuine tripartite steerable assemblages. Our results benefit the distillation of multipartite EPR steering in practice, where the number of copies of initial assemblages is generally finite. 
\end{abstract}

\maketitle
\section{Introduction}
The concept of Einstein-Podolsky-Rosen~(EPR) steering was first noticed by Schrödinger in 1935~\cite{Schrodinger1935discussion}, and describes the phenomena that one partite can remotely steer the state of the other partite by sharing entangled system. Such a quantum feature has been systematically studied in the framework of local hidden state model~\cite{PhysRevLett.98.140402,PhysRevA.76.052116}, which makes its detection quite different from other quantum features, i.e., the detection of entanglement assumes that all the measuring devices used are well characterized~(trusted)~\cite{Horodecki2009RMP}, while the detection of nonlocality depends on device-independent technologies where all the measuring devices used are untrusted~\cite{Brunner2014RMP}. The detection of EPR steering is in the sense that one party uses trusted measurement devices but the other does not, which enables various quantum information processing in semi-device-independent~(SDI) scenarios, such as quantum key distribution~\cite{PhysRevA.85.010301}, randomness certification~\cite{law2014quantum,passaro2015optimal,PhysRevLett.120.260401,coyle2018one}, and channel discrimination~\cite{PhysRevLett.114.060404}. 

The extension of steering to multipartite systems~\cite{RevModPhys.92.015001}, the so-called genuine multipartite EPR steering~\cite{PhysRevA.84.032115,PhysRevLett.111.250403,PhysRevLett.115.010402,cavalcanti2015detection,cavalcanti2016quantum,PhysRevA.98.022101,PhysRevA.97.052307,Lu2020PRL}, is the key resource for quantum information processing in hybrid network where only a few nodes are well characterized and can be trusted. However, the inevitable interactions between a quantum system and its environment can severely degrade the performance of these applications through decoherence. Due to the decoherence, ideal genuine multipartite steering is often not readily shared between remote parties, thus reducing the performance of quantum information tasks. 

In fact, genuine EPR steering, like with some other desirable features of an entangled state such as nonlocality~\cite{PhysRevLett.102.120401,PhysRevA.83.062114,PhysRevA.82.042118,PhysRevLett.106.020402,wu2013hybrid,hoyer2013quantum} and entanglement~\cite{PhysRevA.53.2046,PhysRevLett.80.5239,PhysRevA.60.1888,PhysRevLett.82.1056,horodecki2001distillation}, can be distilled~(boosted) from imperfect multiple copies with local filtering operations~\cite{PhysRevLett.124.120402,PhysRevA.104.022409}. Besides, recent investigations have shown that distillation of steering with local filtering operation is of great interest for quantum foundations as it is closely related to measurement incompatibility~\cite{Ku2022NC,ku2023measurement,hsieh2023characterisation}. It has been shown that at least one copy of a perfectly genuine steerable assemblage can be distilled with certainty from infinite copies of initial assemblages with local filtering operations~\cite{PhysRevA.104.022409}. However, how to maximally enhance the genuine EPR steering from finite number of copies of weakly steerable assemblages has not been extensively studied. 

In this paper, we address this issue by proposing optimal local filtering operation, which is determined by solving the maximization of assemblage fidelity between distilled assemblage and the perfectly genuine steerable assemblage. We derive the analytical expression of filtering operation for $N=2$, and numerical results are presented for the cases of $N>2$. Analytical and numerical results show the advantage of proposed filtering operation as reflected by the enhancement of the tripartite steering in both the one-sided device-independent~(1sDI) scenario and the two-sided device-independent~(2sDI) scenario, especially when $N$ is small and the steerability of initial assemblages is weak. Experimentally, we demonstrate the two-copy distillation with the proposed local filtering operation with optical system. The advantage of the optimal filter is confirmed by the assemblage fidelity and the violation of inequality to witness genuine steerable assemblages, especially when the steerability of initial assemblage is weak.

\section{Tripartite quantum steering}
We start by introducing the scenarios and notations in the detection of tripartite EPR steering~\cite{cavalcanti2015detection}. Considering a tripartite state $\rho^\text{ABC}$ is shared by Alice, Bob and Charlie, there are two SDI scenarios, namely 1sDI scenario and 2sDI scenario, respectively. In 1sDI scenario, Alice's device is uncharacterized, so there is no assumption about Alice’s measurements and the dimension of Alice's subsystem can be arbitrary. Such an unknown measurement can be described by the operator $M_{a|x}$, where the subscript $x\in\mathbb N$ represents the choice of Alice's measurements $A_x$, and the subscript $a\in\mathbb N$ represents the possible outcomes $o_a$. Bob and Charlie's devices are characterized, and they can perform quantum state tomography~(QST) on the qubits in their hands to determine the unnormalized conditional states
\begin{equation}\label{Eq:1sDI_conditional}
\sigma_{a|x}^{\text{BC}}=\Tr_\text{A}(M_{a|x}\otimes\mathds{1}^\text{B}\otimes\mathds{1}^\text{C}\rho^{\text{ABC}}).
\end{equation}
The set $\Sigma_{a|x}=\left\{\sigma_{a|x}^{\text{BC}}\right\}_{a,x}$ is an assemblage. The probability that Alice performs measurement $x$ and obtain the outcome $a$ is $p(a|x)=\Tr(\sigma_{a|x}^{\text{BC}})$, and the normalized quantum state obtained by Bob and Charlie is $\rho^{\text{BC}}_{a|x}=\sigma_{a|x}^{\text{BC}}/p(a|x)$. So the tripartite system is completely described by the conditional distribution $\left\{p(a|x)\right\}_{a,x}$ and normalized states $\left\{\rho^{\text{BC}}_{a|x}\right\}_{a,x}$. 

In 2sDI scenario, Alice's and Bob's measurements are uncharacterized and represented by unknown measurement operators $M_{a|x}$ and $M_{b|y}$ respectively, where the subscripts $y\in\mathbb N$ and $b\in\mathbb N$ represent the choice of Bob's measurements $B_y$ and the possible outcomes $o_b$. Charlie's subsystem is characterized so that QST is performed on Charlie's qubit to determine the unnormalized conditional states
\begin{equation}\label{Eq:2sDI_conditional}
\sigma_{ab|xy}^{\text{C}}=\Tr_\text{AB}(M_{a|x}\otimes M_{b|y}\otimes\mathds{1}^\text{C}\rho^{\text{ABC}}).
\end{equation}
Accordingly, the set $\Sigma_{ab|xy}=\left\{\sigma_{ab|xy}^{\text{C}}\right\}_{a, b, x, y}$ is an assemblage in 2sDI scenario. The probability that Alice and Bob perform the joint measurement $xy$ and obtain the outcome $ab$ is $p(ab|xy)=\Tr(\sigma_{ab|xy}^{\text{C}})$. The normalized state on Charlie's hand is $\rho_{ab|xy}^{\text{C}}=\sigma_{ab|xy}^{\text{C}}/p(ab|xy)$.

For the initial state $\rho^{\text{ABC}}$ does not admit genuine tripartite entanglement, it can be expressed in the form of a mixture of biseparable states in 1sDI~(2sDI) scenario~\cite{cavalcanti2015detection}. If no elements in assemblage $\Sigma_{a|x}$~($\Sigma_{ab|xy}$) can be decomposed into such a mixture of biseparable states, then the assemblage $\Sigma_{a|x}$~($\Sigma_{ab|xy}$) admits genuine tripartite EPR steering. Accordingly, $\rho^{\text{ABC}}$ admits genuine tripartite entanglement in the 1sDI~(2sDI) scenario.   

Consider the tripartite quantum state $\rho^{\text{ABC}}$ that are maximally entangled, such as the Greenberger–Horne–Zeilinger~(GHZ) state $\ket{\text{GHZ}}_3=\frac{1}{\sqrt{2}}(\ket{000}+\ket{111})$, Cavalcanti~\emph{et al.} has proposed inequalities to witness the genuine tripartite EPR steering in two scenarios mentioned above~\cite{cavalcanti2015detection}. In 1sDI scenario, genuine tripartite EPR steering is witnessed by
\begin{equation}\label{Eq:witness_1sDI}
\begin{split}
S_\text{1sDI}=&1+0.1547\langle Z_\text{B}Z_\text{C} \rangle -\frac{1}{3}(\langle A_3Z_\text{B} \rangle +\langle A_3Z_\text{C} \rangle\\
&+\langle A_1X_\text{B}X_\text{C} \rangle-\langle A_1Y_\text{B}Y_\text{C} \rangle -\langle A_2X_\text{B}Y_\text{C} \rangle-\langle A_2Y_\text{B}X_\text{C} \rangle) \geq 0, 
\end{split}
\end{equation}
where $A_x$ for $x\in \left\{1,2,3\right\}$ are the observables with outcomes $a\in\{0, 1\}$ in Alice's uncharacterized measurements and $X$, $Y$ and $Z$ are Pauli matrices. In 2sDI scenario, genuine tripartite EPR steering is witnessed by
\begin{equation}\label{Eq:witness_2sDI}
\begin{split}
S_\text{2sDI}=&1-0.1831( \langle A_3B_3 \rangle +\langle A_3Z_\text{C} \rangle +\langle B_3Z_\text{C} \rangle)-0.2582(\langle A_1B_1X_\text{C} \rangle\\
&-\langle A_1B_2Y_\text{C} \rangle -\langle A_2B_1Y_\text{C} \rangle-\langle A_2B_2X_\text{C} \rangle) \geq 0,
\end{split}
\end{equation}
where $A_x$ and $B_y$ with $x, y\in \left\{1,2,3\right\}$ are the observables with outcomes $a, b\in\{0, 1\}$ in Alice's and Bob’s uncharacterized measurements, respectively. 

In 1sDI scenario, $\ket{\text{GHZ}}_3$ allows the maximum violation of the inequality Eq.~\ref{Eq:witness_1sDI} by $S_\text{1sDI}=-0.8453$ when Alice's measurement is $A_x\in\{X, Y, Z\}$. Correspondingly, the assemblage produced from state $\ket{\text{GHZ}}_3$ with measurements  $A_x\in\{X, Y, Z\}$ is regarded as perfectly genuine tripartite steerable assemblage $\Sigma^\text{GHZ}_{a|x}$. Also, $\ket{\text{GHZ}}_3$ allows the maximum violation of inequality Eq.~\ref{Eq:witness_2sDI} by $S_\text{2sDI}=-0.5820$ when Alice's and Bob's measurements are $A_x, B_y\in\{X, Y, Z\}$, and the assemblage $\Sigma^\text{GHZ}_{ab|xy}$ is called perfectly genuine tripartite steerable assemblage in 2sDI scenario. 

\section{Distillation of tripartite EPR steering with local filtering operation}
The task of genuine steering distillation is to extract $M(M\geq 1)$ copies of perfectly genuine steerable assemblages from $N\geq2 (N>M)$ copies of weakly genuine steerable assemblages using \emph{local filtering} operations, which is free operations that cannot create genuine steerable assemblage from assemblage that do not admit genuine steering~\cite{PhysRevLett.124.120402}. Optimal filtering operation to distill tripartite steering in the case of $N\to\infty$ has been studied theoretically ~\cite{PhysRevA.104.022409}. Here, we focus on distillation of tripartite steering in the cases of finite $N$.

We assume that Bob and Charlie share the assemblage obtained from generalized GHZ~(GGHZ) state
\begin{equation}\label{Eq:GGHZ}
\begin{split}
\ket{\text{GGHZ}}_3=\cos\theta\ket{000} +\sin\theta\ket{111} ,\,\, 0\leq\theta \leq\frac{\pi}{4}. 
\end{split}
\end{equation}
Note that $\ket{\text{GHZ}}_3$ is a special case of GGHZ states with $\theta=\frac{\pi}{4}$. For simplicity, we discuss the details of distillation of assemblage obtained from Eq.~\ref{Eq:GGHZ} in 1sDI scenario. The results of 2sDI scenario is presented and the details can be found in Appendix. 

In 1sDI scenario, the assemblage $\Sigma_{a|x}^\text{GGHZ}=\{\sigma_{a|x}^{\text{BC}}\}_{a, x}$ is obtained from~Eq.~\ref{Eq:GGHZ} when Alice performs $A_0=X$, $A_1=Y$, and $A_2=Z$ measurements. The elements of assemblage $\Sigma_{a|x}^\text{GGHZ}$ are given by 
\begin{equation}\label{Eq:elements_1sDI}
\begin{split}
&\sigma _{0|0}^{\text{BC}}=\frac{1}{2}\ket{\theta _{+}^{0}}\bra{\theta _{+}^{0}}, \sigma _{1|0}^{\text{BC}}=\frac{1}{2}\ket{\theta _{-}^{0}}\bra{\theta _{-}^{0}},
\\
&\sigma _{0|1}^{\text{BC}}=\frac{1}{2}\ket{\theta _{-}^{1}}\bra{\theta _{-}^{1}}, \sigma _{1|1}^{\text{BC}}=\frac{1}{2}\ket{\theta _{+}^{1}}\bra{\theta _{+}^{1}},
\\
&\sigma _{0|2}^{\text{BC}}=\cos^2\theta\ket{00}\bra{00},\sigma _{1|2}^{\text{BC}}=\sin^2\theta\ket{11}\bra{11},
\end{split}
\end{equation}
where $\ket{\theta _{\pm}^{0}}=\cos\theta\ket{00}\pm\sin\theta\ket{11}$ and $\ket{\theta _{\pm}^{1}}=\cos\theta\ket{00}\pm i\sin\theta\ket{11}$. Note that the assemblage $\Sigma_{a|x}^\text{GGHZ}$ cannot reach the maximum violation of inequality~Eq.~\ref{Eq:witness_1sDI} so that $\Sigma_{a|x}^\text{GGHZ}$ is considered to be a weakly steerable assemblage. Furthermore, $S_\text{1sDI}$ is a monotonic function of $\theta$ within the range of $0<\theta<\frac{\pi}{4}$, and $\Sigma_{a|x}^\text{GGHZ}$ violates inequality~Eq.~\ref{Eq:witness_1sDI} for $\theta \in \left( 0.185,\frac{\pi}{4} \right]$. 

In distillation protocol, only one trusted party, say Charlie, performs local filtering operations in 1sDI and 2sDI scenarios:
\begin{itemize}
\item[(1)] First, Charlie perform a dichotomic POVM $\{C_0^\dagger(\kappa)C_0(\kappa), C_1^\dagger(\kappa)C_1(\kappa)\}$ with
\begin{equation}\label{Eq:filter}
C_0(\kappa)=\kappa\ket{0}\bra{0}+\ket{1}\bra{1}, C_1(\kappa)=\sqrt{1-\kappa^2}\ket{0}\bra{0},
\end{equation} 
satisfying $C_{0(1)}^\dagger(\kappa)C_{0(1)}(\kappa)\geq0$ and $C_0^\dagger(\kappa)C_0(\kappa)+C_1^\dagger(\kappa)C_1(\kappa)=\mathds 1$, on the $n$th~($n\in\{1,2,\cdots,N-1\}$) copy of $\Sigma_{a|x}^\text{GGHZ}$. Hereafter, $C_0(\kappa)$ is referred as filtering operation and denoted as $C_\text{F}(\kappa)$. Accordingly, the POVM with outcome $c_1=0$ indicates success of filtering operation, while POVM with outcome $c_1=1$ indicates the failure of filtering operation. The output of the POVM is denoted as a bit string $\{c_1, c_2, \cdots, c_{N-1}\}$ 
\item[(2)] Charlie sets $c_{N}=1$ for the $N$th copy if he gets the output $c_n=0$ for $n\in\{1,2,\cdots,N-1\}$, otherwise sets $c_{N}=0$ without measuring $N$th copy.

\item[(3)] Charlie sends the bit string $\bm{c}=\left\{c_{1}, c_{2},\cdots, c_N \right\}$ to Alice and Bob. All parties discard every $n$th copy for which $c_n=1$. The output of this protocol is the remaining assemblages $\Sigma_{a|x}^\text{dist}=\{\sigma_{a|x}^{\text{dist}}\}_{a, x}$. 
\end{itemize}
Conditional upon a successful filtering operation on $n$th copy, which occurs with probability
\begin{equation}
P_\text{succ}=\Tr\left[(\mathds 1\otimes C_\text{F}(\kappa))\rho^{\text{BC}}(\mathds 1\otimes C_\text{F}(\kappa)^\dagger)\right]=\kappa^2\cos^2\theta+\sin^2\theta,
\end{equation}
the assemblage $\Sigma_{a|x}^\text{GGHZ}$ is updated to $\tilde{\Sigma}_{a|x}=\{\tilde{\sigma}_{a|x}^\text{BC}\}_{a,x}$ with 
\begin{equation}
\tilde{\sigma}_{a|x}^\text{BC}=\frac{1}{P_\text{succ}}(\mathds 1\otimes C_\text{F}(\kappa))\sigma^{\text{BC}}_{a|x}(\mathds 1\otimes C_\text{F}(\kappa)^\dagger).
\end{equation}

The probability that filtering operations are failed for all $N-1$ copies is $P^{\text{1sDI}, N}_\text{fail}=(1-P_\text{succ})^{N-1}$. Consequently, the probability that at least one assemblage $\tilde{\Sigma}_{a|x}$ can be distilled from $N-1$ copies is $P^{\text{1sDI}, N}_\text{succ}=1-(1-P_\text{succ})^{N-1}$. Thus, the output assemblage according to distillation scheme is 
\begin{equation}\label{Eq:distassN}
\Sigma_{a|x}^{\text{dist}, N}=P^{\text{1sDI}, N}_\text{succ}\tilde{\Sigma}_{a|x}+P^{\text{1sDI}, N}_\text{fail}\Sigma_{a|x}^\text{GGHZ}.
\end{equation}  

The figure of merit to determine optimal $C_\text{F}(\kappa)$ is assemblage fidelity~\cite{PhysRevLett.124.120402} between the distilled assemblage $\Sigma_{a|x}^\text{dist}$ and the perfectly steerable assemblage $\Sigma_{a|x}^\text{GHZ}$
\begin{equation}\label{Eq:Fidelity_1sDI}
F _\text{1sDI}\left(\Sigma_{a|x}^{\text{dist}, N},\Sigma_{a|x}^\text{GHZ} \right) =\underset{x}{\min}\sum_{a}{f \left( \sigma _{a|x}^{\text{dist}, N},\sigma _{a|x}^\text{GHZ} \right)}
\end{equation}
where $f \left( \sigma, \rho \right) =\Tr\left[ \sqrt{\sqrt{\sigma}\rho\sqrt{\sigma}} \right]$.  Note that $F_\text{1sDI} \leq1$ with equality holds if $\Sigma_{a|x}^{\text{dist}, N}=\Sigma_{a|x}^\text{GHZ}$. Thus, the optimal local filtering operation is determined by solving the maximization of
\begin{equation}
\label{Eq:opt}
    \begin{split}
        \text{maximize}\hspace{1.2cm}&F _\text{1sDI}\left(\Sigma_{a|x}^{\text{dist}, N},\Sigma_{a|x}^\text{GHZ} \right)\\
        \text{subject to}\hspace{1.2cm}&0\leq\kappa\leq1,
    \end{split}
\end{equation}   

\subsection{Optimal local filtering operation in two-copy distillation}
We start with the simplest case of $N=2$. According to Eq.~\ref{Eq:distassN}, distilled assemblage is
\begin{equation}\label{Eq:distass2}
\begin{split}
\Sigma_{a|x}^{\text{dist}, 2}&=P_\text{succ}\tilde{\Sigma}_{a|x}+(1-P_\text{succ})\Sigma_{a|x}^\text{GGHZ}\\
&=(\kappa^2\cos^2\theta+\sin^2\theta)\tilde{\Sigma}_{a|x}+\left(1-\kappa^2\right)\cos^2\theta\Sigma_{a|x}^\text{GGHZ}.
\end{split}
\end{equation}
Thus, $F_\text{1sDI}$ is 
\begin{equation}
\begin{split}
F _\text{1sDI}\left(\Sigma_{a|x}^{\text{dist}, 2},\Sigma_{a|x}^\text{GHZ} \right)&=\underset{x}{\min}\sum_{a}{f \left( \sigma _{a|x}^\text{dist},\sigma _{a|x}^\text{GHZ} \right)}\\
&=\sum_{a=0, 1} f(\sigma_{a|0}^\text{dist}, \sigma_{a|0}^\text{GHZ})\\
&=\sqrt{\frac{1}{2}+\cos\theta\sin\theta(\cos^2\theta-\kappa^2\cos^2\theta+\kappa)}\\
&\leq\sqrt{\frac{1}{2}+\cos\theta\sin\theta\left(\cos^2\theta+\frac{1}{4\cos^2\theta}\right)}. 
\end{split}
\end{equation}
The equality holds when $\kappa=\frac{1}{2\cos^2\theta}$, and the corresponding
\begin{equation}
C_\text{F}(\kappa)=\frac{1}{2\cos^2\theta}\ket{0}\bra{0}+\ket{1}\bra{1}, 
\end{equation}
is the optimal local filtering operation, yielding the maximal assemblage fidelity
\begin{equation}
F_\text{1sDI}^\kappa=\sqrt{\frac{1}{2}+\cos\theta\sin\theta\left(\cos^2\theta+\frac{1}{4\cos^2\theta}\right)}
\end{equation}
The detailed derivation can be found in Appendix. 

\begin{figure}[h!t]
\centering
	\includegraphics[width=0.85\linewidth]{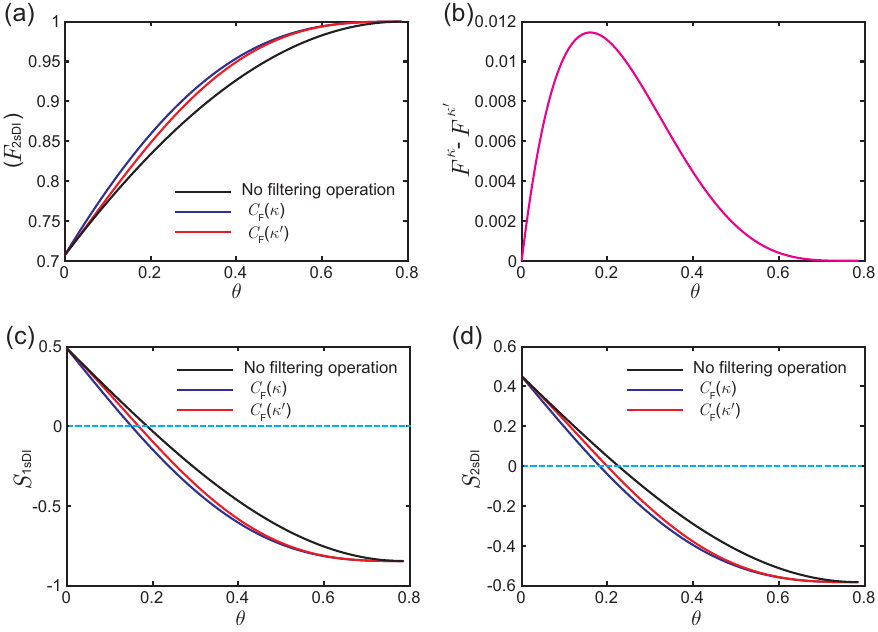}
\caption{Theoretical results of steering distillation with filtering operation $C_\text{F}(\kappa)$ and $C_\text{F}(\kappa^\prime)$. (a), $F_\text{1sDI}$ ($F_\text{2sDI}$). (b), $F^{\kappa}_\text{1sDI}-F^{\kappa^\prime}_\text{1sDI}$ ($F^{\kappa}_\text{2sDI}-F^{\kappa^\prime}_\text{2sDI}$). (c), $S_\text{1sDI}$. (d), $S_\text{2sDI}$. Black lines: no filtering operation is performed. Blue lines: filtering operation $C_\text{F}(\kappa)$ is performed. Red lines: filtering operation $C_\text{F}(\kappa^\prime)$ is performed.}
\label{Fig:theory}
\end{figure}

To give a comparison, we consider the local filtering operation $C_\text{F}(\kappa^\prime)$ with $\kappa^\prime=\tan\theta$~\cite{PhysRevA.104.022409}. Note that $C_\text{F}(\kappa^\prime)$ is optimal in the regime of infinite copies~($N\to\infty$) of $\Sigma_{a|x}^\text{GGHZ}$~(see Appendix for derivation). In two-copy scenario, maximal assemblage fidelity with $C_\text{F}(\kappa^\prime)$ is 
\begin{equation}\label{Eq:Finfty}
F^{\kappa^\prime}_\text{1sDI}=\sqrt{1-\frac{1}{2}(1-\sin2\theta)\cos2\theta}.
\end{equation}

The comparison of assemblage fidelity with local filtering operation $C_\text{F}(\kappa)$ and $C_\text{F}(\kappa^\prime)$ is shown in Figure~\ref{Fig:theory}(a), which clearly indicates that both $C_\text{F}(\kappa)$ and $C_\text{F}(\kappa^\prime)$ can enhance steerability. Compared to $C_\text{F}(\kappa^\prime)$, the maximum enhancement using $C_\text{F}(\kappa)$ is about 0.012 at $\theta=0.18$ as shown in~Figure~\ref{Fig:theory}(b). More importantly, for assemblages $\Sigma_{a|x}^\text{GHZ}$ with $\theta\in(0.151,0.185]$ that do not admits genuine tripartite EPR steering according to~Eq.~\ref{Eq:witness_1sDI}, $C_\text{F}(\kappa)$ activates them to be steerable assemblages as shown in~Figure~\ref{Fig:theory}(c). Similar phenomena also exist in 2sDI scenario as shown in~Figure~\ref{Fig:theory}(d).

\begin{figure}[h!t]
\centering
	\includegraphics[width=\linewidth]{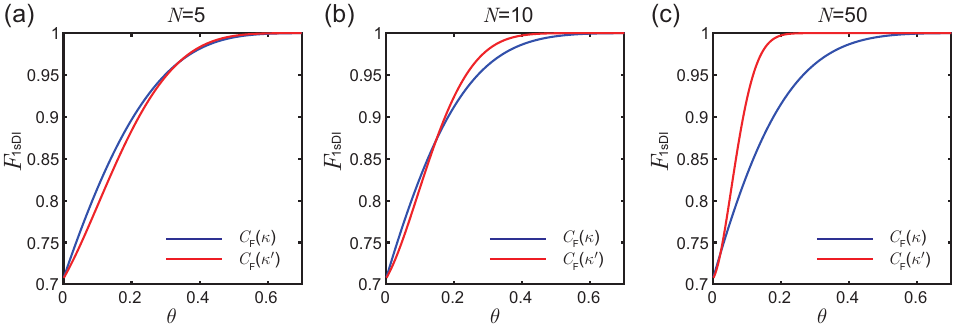}
\caption{Theoretical results of assemblage fidelity $F_\text{1sDI}\left(\Sigma_{a|x}^{\text{dist},N},\Sigma_{a|x}^\text{GHZ}\right)$ in $N$-copy 1sDI distillation scenario with (a) $N=5$, (b) $N=10$ and (c) $N=50$. The blue and red lines represents the results with local filtering operations $C_\text{F}(\kappa)$ and $C_\text{F}(\kappa^\prime)$, respectively.}
\label{Fig:theory_com}
\end{figure}

We also investigate the performance of filtering operations $C_\text{F}(\kappa)$ and $C_\text{F}(\kappa^\prime)$ in 1sDI scenario with $N>2$ copies. We calculate the assemblage fidelity $F_\text{1sDI}\left(\Sigma_{a|x}^{\text{dist}, N},\Sigma_{a|x}^\text{GHZ}\right)$ for $N=5, 10$ and $50$ respectively, and the results are shown in Figure~\ref{Fig:theory_com}. It is evident that filtering operation $C_\text{F}(\kappa)$ outperforms $C_\text{F}(\kappa^\prime)$ for smaller $N$ and $\theta$. In the case of larger $N$, the successful probability $P^{\text{1sDI}, N}_\text{succ}$ gets closer to 1, and $C_\text{F}(\kappa^\prime)$ exhibits better performance as the target assemblage with $C_\text{F}(\kappa^\prime)$ is $\tilde{\Sigma}_{a|x}= \Sigma_{a|x}^\text{GHZ}$~\cite{PhysRevA.104.022409}.

\subsection{Optimal local filtering operation in $N$-copy distillation}
\begin{figure}[h!t]
\centering
\includegraphics[width=\linewidth]{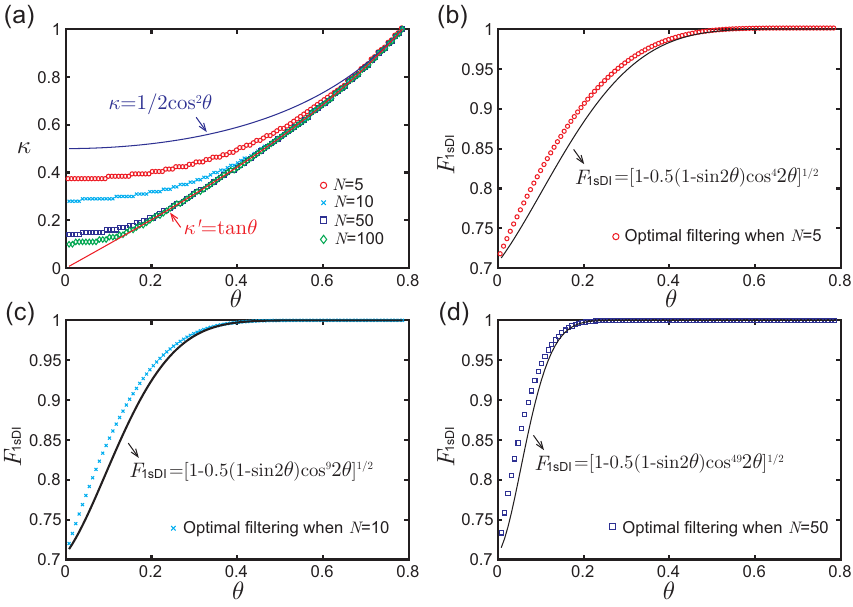}
\caption{Numerical results of optimal local filtering operation in $N$-copy 1sDI distillation. (a), the optimal values of $\kappa$ with $N=5$~(red circle), 10~(cyan cross), 50~(blue square) and 100~(greed diamond). The blue solid line represents the optimal value of $\kappa^\prime=\frac{1}{2\cos^2\theta}$ with $N=2$, while the red solid line represents the optimal value of $\kappa^\prime=\tan\theta$ with $N\to\infty$. The assemblage fidelity $F_\text{1sDI}$ with optimal $C_\text{F}(\kappa)$ in 5-copy distillation~(b), 10-copy distillation~(c) and 50-copy distillation~(d). The black sold lines in (b), (c) and (d) represent the analytic expression of $F_\text{1sDI}=\sqrt{1-\frac{1}{2}(1-\sin2\theta)\cos^{N-1}2\theta}$ with local filtering operation $C_\text{F}(\kappa^\prime)$. }
\label{Fig:filterNcopy}
\end{figure}

In the regime of distillation from $N$ copies initial assemblages, it is complicated to derive the analytic expression of optimal local filtering operation via maximization of $F_\text{1sDI}\left(\Sigma_{a|x}^{\text{dist}, N},\Sigma_{a|x}^\text{GHZ}\right)$. The optimal filter operation can be determined numerically. The numerical results of optimal filtering operation of $N$-copy distillation with $N=5, 10, 50$ and $100$ are shown in Figure~\ref{Fig:filterNcopy}(a), where the optimal value $\kappa$ converges to $\kappa^\prime$ as in increases. In For $N=5, 10$ and $50$, we calculate the assemblage fidelity $F_\text{1sDI}$ with optimal local filtering operation derived in Figure~\ref{Fig:filterNcopy}(a),  and the results are shown in Figure~\ref{Fig:filterNcopy}(b), (c) and (d) respectively. Clearly, $N$-copy distillation with $C_\text{F}(\kappa)$ enhances the assemblage fidelity compared to that with $C_\text{F}(\kappa^\prime)$.   

\section{Experimental demonstration}
\begin{figure}[h!tbp]
\centering
\includegraphics[width=0.8\linewidth]{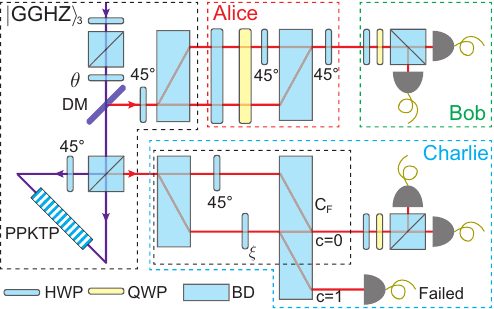}
\caption{Schematic drawing of experimental setup to investigate the two-copy distillation protocol.}
\label{Fig:setup}
\end{figure}

We experimentally demonstrate the distillation of genuine tripartite quantum steering with optical system. The photon pairs are generated on a potassium titanium phosphate~(PPKTP) crystal via spontaneous parametric down conversion~(SPDC).  As shown in Figure~\ref{Fig:setup}, we first generate a pair of polarization-entangled photons by bidirectionally pumping a PPKTP crystal set at the Sagnac interferometer. Here, the pump light has a central wavelength of 405~nm and the photons generated from the PPKTP have a central wavelength of 810~nm. The polarization-entangled photons are with ideal form of $\cos\theta\ket{HV}+\sin\theta\ket{VH}$ where $H$ and $V$ denote the horizontal and vertical polarization respectively. The parameter $\theta$ is determined by the polarization of pump light. One photon passes through an half-wave plate~(HWP) set at 45$^\circ$, followed by a beam displacer that transmits vertical polarization and deviates horizontal polarization. This produces the hybrid-coded three-qubit GGHZ states $\ket{\text{GGHZ}}_3$
\begin{equation}
\ket{\text{GGHZ}}_3=\cos\theta\ket{HhH}_\text{ABC}+\sin\theta\ket{VvV}_\text{ABC}, 
\end{equation} 
where $h$ and $v$ are the deviated and transmitted modes, respectively. Specifically, Alice is encoded in polarization degrees of freedom~(DOF), Bob is encoded in path DOF and Charlie is encoded in polarization DOF. The projective measurement can be performed on each party individually~\cite{PhysRevResearch.3.023228, PhysRevLett.127.200501}. 

The full demonstration of two-copy distillation protocol requires two copies of $\ket{\text{GGHZ}}_3$. However, the distillation does not require joint quantum operation between qubits form individual copy, so that it can be equivalently realized by running the experiments with two different experimental settings and post-processing for a proof-of-principle demonstration. Specifically, the first and second experimental setting corresponds to the first and second copy respectively, and the classical communications are simulated with pot-processing of collected data. In the first experimental setting, Charlie performs the filtering operation $C_\text{F}(\kappa)$, which is realized with two HWPs and two BDs as shown in Figure~\ref{Fig:setup}. The parameter $\kappa$ is determined by the angle $\xi$ of HWP, i.e., $\xi=\text{arc}\sin\kappa/2$. If filtering operation succeeds~(fails), the photon would come out from the upper~(lower) port. Charlie then records the photons coming out from port $c=0$ and $c=1$, and calculates the probability of $P_\text{succ}^\text{1sDI}$. For the successfully filtered state, Alice performs measurements of $A_x\in\{X, Y, Z\}$ and records the probability of outcomes $p_{a|x}$. Bob and Charlie reconstruct $\rho^\text{BC}_{a|x}$ and then obtain the assemblage of $\tilde{\Sigma}_{a|x}$. In the second experiment setting, Charlie sets $\xi=45^\circ$, which corresponds to identity operation. Alice, Bob and Charlie perform the same measurements as in the first experiment and then obtain the assemblage of $\Sigma_{a|x}^\text{GGHZ}$. The filtering operation $C_\text{F}(\kappa)$ is realized with two HWPs and two BDs as shown in Figure~\ref{Fig:setup}. The parameter $\kappa$ is determined by the angle $\xi$ of HWP, i.e., $\xi=\text{arc}\sin\kappa/2$. If filtering operation succeeds~(fails), the photon would come out from the upper~(lower) port.

\begin{figure}[ht]
\centering
	\includegraphics[width=0.85\linewidth]{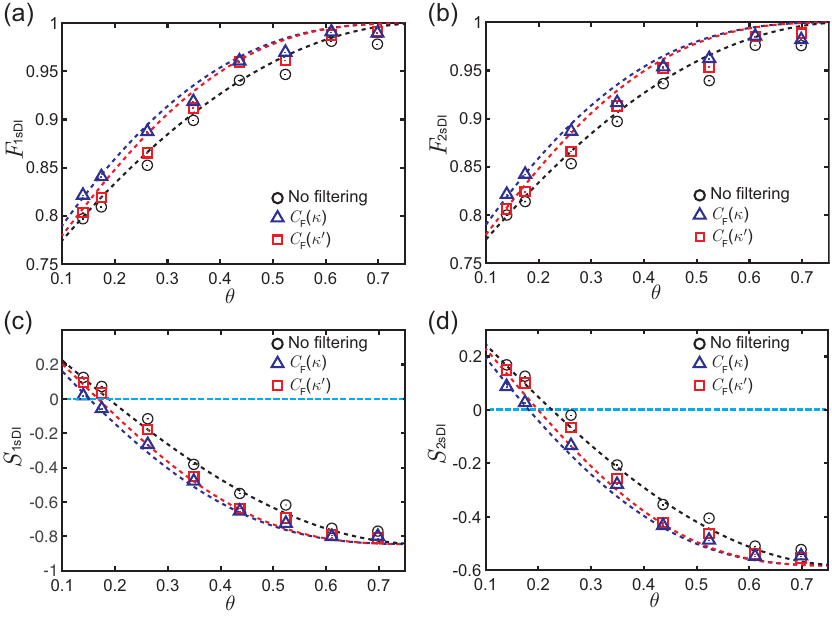}
\caption{Experimental results of $F_\text{1sDI}$ (a), $F_\text{2sDI}$ (b), $S_\text{1sDI}$ (c) and $S_\text{ssDI}$ (d). Black circles: no filtering operation is performed. Blue triangles: filtering operation $C_\text{F}(\kappa)$ is performed. Red squares: filtering operation $C_\text{F}(\kappa^\prime)$ is performed. The dashed lines represents the corresponding theoretical predictions. The error bars are too small compared to size of markers. }
\label{Fig:expdata}
\end{figure}

With such an experimental setting and data collection, we can calculate the distilled assemblage $\Sigma_{a|x}^\text{dist}$ according to Eq.~\ref{Eq:distass2}. The average assemblage $\Sigma_{ab|xy}^\text{dist}$ in 2sDI scenario is obtained using the same approach. In our experiment, we prepare eight GGHZ states $\ket{\text{GGHZ}}_3$ with $\theta\in[\frac{\pi}{50}, \frac{\pi}{18}, \frac{\pi}{12}, \frac{\pi}{8}, \frac{5\pi}{36}, \frac{\pi}{6}, \frac{7\pi}{36}, \frac{2\pi}{9}]$. For each state, we perform the distillation with filtering operation $C_\text{F}(\kappa)$ and $C_\text{F}(\kappa^\prime)$, and calculate the assemblage fidelities $F_\text{1sDI}(\Sigma_{a|x}^\text{dist}, \Sigma_{a|x}^\text{GHZ})$ and $F_\text{2sDI}(\Sigma_{ab|xy}^\text{dist}, \Sigma_{ab|xy}^\text{GHZ})$. The results are shown with blue triangles and red squares in~Figure~\ref{Fig:expdata}(a) and~Figure~\ref{Fig:expdata} (b), respectively. We observe both $C_\text{F}(\kappa)$ and $C_\text{F} (\kappa^\prime)$ can improve the assemblage fidelity. In particular, $C_\text{F}(\kappa)$ outperforms  $C_\text{F}(\kappa^\prime)$ for initial assemblages $\Sigma^\text{GGHZ}$ with smaller $\theta$~(weaker steerability).

Furthermore, we detect the EPR steering witnesses in Eq.~\ref{Eq:witness_1sDI} and Eq.~\ref{Eq:witness_2sDI}, and the results are shown in~Figure~\ref{Fig:expdata}(c) and~Figure~\ref{Fig:expdata} (d) respectively. In 1sDI scenario, we observe that $\ket{\text{GGHZ}}_3$ with $\theta=\pi/50$ and $\theta=\pi/18$ cannot violate Eq.~\ref{Eq:witness_1sDI}. For $\theta=\pi/18$, the distilled assemblage with filtering operation $C_\text{F}(\kappa)$ admits genuine EPR steering while that with filtering operation $C_\text{F}(\kappa^\prime)$ does not. Note that there are discrepancies between experimental results and theoretical predictions as shown in Figure~\ref{Fig:expdata}. This is mainly caused by the experimental imperfections in state preparation and manipulation, including higher-order emissions in SPDC, mode mismatch when overlapping two photons in Sagnac interferometer and the accuracy of waveplates. For the noisy state $\rho_\text{noisy}$, the corresponding optimal filtering operation can be determined by the maximization of assemblage fidelity aforementioned. 

\section{Conclusion}
In conclusion, we investigate the distillation of genuine tripartite steerable assemblage from $N$ weakly steerable assemblages using a local filtering operation, in both the 1sDI and 2sDI scenarios. We propose the optimal local filtering operation that maximally enhances the assemblage fidelity of distilled assemblage in $N$-copy distillation scenarios. Experimentally, we perform a proof-of-principle demonstration of the proposed distillation scheme with optical system. The experimental results convince the theoretical predictions, and show advantages over other filtering operation in practice.   

\section*{Acknowledgment}

This work is supported by the National Key R\&D Program of China (Grant No.
2019YFA0308200), the National Natural Science Foundation of China (Grants No. 11974213 and No. 92065112), Shenzhen Fundamental Research Program (Grants No. JCYJ20190806155211142 and No. JCYJ20220530141013029), Shandong Provincial Natural Science Foundation (Grant No. ZR2020JQ05), Taishan Scholar of Shandong Province (Grant No. tsqn202103013), Shandong University Multidisciplinary Research and Innovation Team of Young Scholars (Grant No. 2020QNQT), and the Higher Education Discipline Innovation Project (‘111’) (Grant No. B13029).

\section*{Disclosures}
 The authors declare no conflicts of interest.

\section*{Data availability}
Data underlying the results presented in this paper are not publicly available at this time but may be obtained from the authors upon reasonable request.
\bibliography{steering}
\appendix
\onecolumngrid
\section{The derivation of $\kappa$ in 1sDI scenario}
\subsection{Explicit form of assemblage $\sum_{a|x}^\text{dist}$}
\begin{equation}
\begin{split}
\sigma_{0|0}^\text{dist}&=P_\text{succ}\tilde{\sigma}_{0|0}+P_\text{fail}\sigma_{0|0}^\text{GGHZ}\\
&=\frac{1}{2}\begin{pmatrix}
\kappa^2\cos^2\theta & 0 &0 & \kappa\cos\theta\sin\theta \\
0&0&0&0\\
0&0&0&0\\
\kappa\cos\theta\sin\theta & 0 &0&\sin^2\theta
\end{pmatrix}+\frac{(1-\kappa^2)\cos^2\theta}{2}\begin{pmatrix}
\cos^2\theta & 0 &0& \cos\theta\sin\theta \\
0&0&0&0\\
0&0&0&0\\
\cos\theta\sin\theta & 0 &0&\sin^2\theta
\end{pmatrix}\\
&=\frac{1}{2}\begin{pmatrix}
\kappa^2\cos^2\theta\sin^2\theta+\cos^4\theta & 0 &0 & \cos\theta\sin\theta(\kappa+\cos^2\theta-\kappa^2\cos^2\theta) \\
0&0&0&0\\
0&0&0&0\\
\cos\theta\sin\theta(\kappa+\cos^2\theta-\kappa^2\cos^2\theta) & 0 &0&\sin^2\theta(1+\cos^2\theta-\kappa^2\cos^2\theta)
\end{pmatrix}\\
\end{split}
\end{equation}

\begin{equation}
\begin{split}
\sigma_{1|0}^\text{dist}&=P_\text{succ}\tilde{\sigma}_{1|0}+P_\text{fail}\sigma_{1|0}^\text{GGHZ}\\
&=\frac{1}{2}\begin{pmatrix}
\kappa^2\cos^2\theta & 0 &0 & -\kappa\cos\theta\sin\theta \\
0&0&0&0\\
0&0&0&0\\
-\kappa\cos\theta\sin\theta & 0 &0&\sin^2\theta
\end{pmatrix}+\frac{(1-\kappa^2)\cos^2\theta}{2}\begin{pmatrix}
\cos^2\theta & 0 &0& -\cos\theta\sin\theta \\
0&0&0&0\\
0&0&0&0\\
-\cos\theta\sin\theta & 0 &0&\sin^2\theta
\end{pmatrix}\\
&=\frac{1}{2}\begin{pmatrix}
\kappa^2\cos^2\theta\sin^2\theta+\cos^4\theta & 0 &0 & -\cos\theta\sin\theta(\kappa+\cos^2\theta-\kappa^2\cos^2\theta) \\
0&0&0&0\\
0&0&0&0\\
-\cos\theta\sin\theta(\kappa+\cos^2\theta-\kappa^2\cos^2\theta) & 0 &0&\sin^2\theta(1+\cos^2\theta-\kappa^2\cos^2\theta)
\end{pmatrix}\\
\end{split}
\end{equation}

\begin{equation}
\begin{split}
\sigma_{0|1}^\text{dist}&=P_\text{succ}\tilde{\sigma}_{0|1}+P_\text{fail}\sigma_{0|1}^\text{GGHZ}\\
&=\frac{1}{2}\begin{pmatrix}
\kappa^2\cos^2\theta & 0 &0 & -i\kappa\cos\theta\sin\theta \\
0&0&0&0\\
0&0&0&0\\
i\kappa\cos\theta\sin\theta & 0 &0&\sin^2\theta
\end{pmatrix}+\frac{(1-\kappa^2)\cos^2\theta}{2}\begin{pmatrix}
\cos^2\theta & 0 &0& -i\cos\theta\sin\theta \\
0&0&0&0\\
0&0&0&0\\
i\cos\theta\sin\theta & 0 &0&\sin^2\theta
\end{pmatrix}\\
&=\frac{1}{2}\begin{pmatrix}
\kappa^2\cos^2\theta\sin^2\theta+\cos^4\theta & 0 &0 & -i\cos\theta\sin\theta(\kappa+\cos^2\theta-\kappa^2\cos^2\theta) \\
0&0&0&0\\
0&0&0&0\\
i\cos\theta\sin\theta(\kappa+\cos^2\theta-\kappa^2\cos^2\theta) & 0 &0&\sin^2\theta(1+\cos^2\theta-\kappa^2\cos^2\theta)
\end{pmatrix}\\
\end{split}
\end{equation}

\begin{equation}
\begin{split}
\sigma_{1|1}^\text{dist}&=P_\text{succ}\tilde{\sigma}_{1|1}+P_\text{fail}\sigma_{1|1}^\text{GGHZ}\\
&=\frac{1}{2}\begin{pmatrix}
\kappa^2\cos^2\theta & 0 &0 & i\kappa\cos\theta\sin\theta \\
0&0&0&0\\
0&0&0&0\\
-i\kappa\cos\theta\sin\theta & 0 &0&\sin^2\theta
\end{pmatrix}+\frac{(1-\kappa^2)\cos^2\theta}{2}\begin{pmatrix}
\cos^2\theta & 0 &0& i\cos\theta\sin\theta \\
0&0&0&0\\
0&0&0&0\\
-i\cos\theta\sin\theta & 0 &0&\sin^2\theta
\end{pmatrix}\\
&=\frac{1}{2}\begin{pmatrix}
\kappa^2\cos^2\theta\sin^2\theta+\cos^4\theta & 0 &0 & i\cos\theta\sin\theta(\kappa+\cos^2\theta-\kappa^2\cos^2\theta) \\
0&0&0&0\\
0&0&0&0\\
-i\cos\theta\sin\theta(\kappa+\cos^2\theta-\kappa^2\cos^2\theta) & 0 &0&\sin^2\theta(1+\cos^2\theta-\kappa^2\cos^2\theta)
\end{pmatrix}\\
\end{split}
\end{equation}

\begin{equation}
\begin{split}
\sigma_{0|2}^\text{dist}&=P_\text{succ}\tilde{\sigma}_{0|2}+P_\text{fail}\sigma_{0|2}^\text{GGHZ}\\
&=\begin{pmatrix}
\kappa^2\cos^2\theta & 0 &0 & 0 \\
0&0&0&0\\
0&0&0&0\\
0 & 0 &0&0
\end{pmatrix}+(1-\kappa^2)\cos^2\theta\begin{pmatrix}
\cos^2\theta & 0 &0& 0 \\
0&0&0&0\\
0&0&0&0\\
0 & 0 &0&0
\end{pmatrix}=\begin{pmatrix}
\kappa^2\cos^2\theta\sin^2\theta+\cos^4\theta & 0 &0 & 0 \\
0&0&0&0\\
0&0&0&0\\
0 & 0 &0&0
\end{pmatrix}\\
\end{split}
\end{equation}

\begin{equation}
\begin{split}
\sigma_{1|2}^\text{dist}&=P_\text{succ}\tilde{\sigma}_{1|2}+P_\text{fail}\sigma_{1|2}^\text{GGHZ}\\
&=\begin{pmatrix}
0 & 0 &0 & 0 \\
0&0&0&0\\
0&0&0&0\\
0 & 0 &0&\sin^2\theta
\end{pmatrix}+(1-\kappa^2)\cos^2\theta\begin{pmatrix}
0 & 0 &0& 0 \\
0&0&0&0\\
0&0&0&0\\
0 & 0 &0&\sin^2\theta
\end{pmatrix}=\begin{pmatrix}
0 & 0 &0 & 0 \\
0&0&0&0\\
0&0&0&0\\
0 & 0 &0&\sin^2\theta(1+\cos^2\theta-\kappa^2\cos^2\theta)
\end{pmatrix}\\
\end{split}
\end{equation}

\subsection{Maximization of assemblage fidelity}
The assemblage fidelity is 
\begin{equation}
\begin{split}
\sum_{a=0, 1} f(\sigma_{a|0}^\text{dist}, \sigma_{a|0}^\text{GHZ})=\sum_{a=0, 1} f(\sigma_{a|1}^\text{dist}, \sigma_{a|1}^\text{GHZ})=\sqrt{\frac{1}{2}+\cos\theta\sin\theta(\cos^2\theta-\kappa^2\cos^2\theta+\kappa)}\\
\sum_{a=0, 1} f(\sigma_{a|2}^\text{dist}, \sigma_{a|2}^\text{GHZ})=\sqrt{\frac{\kappa^2\cos^2\theta\sin^2\theta+\cos^4\theta}{2}}+\sqrt{\frac{\sin^2\theta(1+\cos^2\theta-\kappa^2\cos^2\theta)}{2}}
\end{split}
\end{equation}
It is easy to check 
\begin{equation}
\begin{split}
&\left[\sum_{a=0, 1} f(\sigma_{a|0}^\text{dist}, \sigma_{a|0}^\text{GHZ})\right]^2-\left[\sum_{a=0, 1} f(\sigma_{a|2}^\text{dist}, \sigma_{a|2}^\text{GHZ})\right]^2\\
=&\cos\theta\sin\theta\left[(1-\kappa^2)\cos^2\theta+\kappa-\sqrt{(1-\kappa^2)^2\cos^4\theta+(1+\kappa^2)(1-\kappa^2)\cos^2\theta+\kappa^2}\right]\\
\leq&\cos\theta\sin\theta\left[(1-\kappa^2)\cos^2\theta+\kappa-\sqrt{(1-\kappa^2)^2\cos^4\theta+2\kappa(1-\kappa^2)\cos^2\theta+\kappa^2}\right]\\
=&0,
\end{split}
\end{equation}
and we have the assemblage fidelity of 
\begin{equation}
F_\text{1sDI}=\sqrt{\frac{1}{2}+\cos\theta\sin\theta(\cos^2\theta-\kappa^2\cos^2\theta+\kappa)}\leq\sqrt{\frac{1}{2}+\cos\theta\sin\theta\left(\cos^2\theta+\frac{1}{4\cos^2\theta}\right)},
\end{equation}
with the equality holding when $\kappa=\frac{1}{2\cos^2\theta}$.

\section{The derivation of $\kappa$ in 2sDI scenario}
\subsection{Explicit form of assemblage $\Sigma_{ab|xy}^\text{GGHZ}$}
The components of assemblages $\Sigma_{ab|xy}^\text{GGHZ}$ are given by
\begin{equation}\label{eq18}
\begin{split}
&\sigma _{00|00}^\text{C}=\sigma _{11|00}^\text{C}=\sigma _{01|11}^\text{C}=\sigma _{10|11}^\text{C}=\frac{1}{4}\ket{\theta_{+}^{2}}\bra{\theta_{+}^{2}},
\sigma _{01|00}^\text{C}=\sigma _{10|00}^\text{C}=\sigma _{00|11}^\text{C}=\sigma _{11|11}^\text{C}=\frac{1}{4}\ket{\theta_{-}^{2}}\bra{\theta_{-}^{2}},
\\
&\sigma _{00|01}^\text{C}=\sigma _{11|01}^\text{C}=\sigma _{00|10}^\text{C}=\sigma_{11|10}^\text{C}=\frac{1}{4}\ket{\theta_{-}^{3}}\bra{\theta_{-}^{3}},
\sigma _{01|01}^\text{C}=\sigma _{10|01}^\text{C}=\sigma _{01|10}^\text{C}=\sigma _{10|10}^\text{C}=\frac{1}{4}\ket{\theta_{+}^{3}}\bra{\theta_{+}^{3}},
\\
&\sigma _{00|02}^\text{C}=\sigma _{10|02}^\text{C}=\sigma_{00|12}^\text{C}=\sigma_{10|12}^\text{C}=\sigma_{00|20}^\text{C}
=\sigma _{01|20}^\text{C}=\sigma _{00|21}^\text{C}=\sigma _{01|21}^\text{C}=\frac{\cos^2\theta}{2}\ket{0}\bra{0},
\\
&\sigma _{01|02}^\text{C}=\sigma _{11|02}^\text{C}=\sigma _{01|12}^\text{C}=\sigma _{11|12}^\text{C}=\sigma _{10|20}^\text{C}=\sigma _{11|20}^\text{C}=\sigma _{10|21}^\text{C}=\sigma _{11|21}^\text{C}=\frac{\sin^2\theta}{2}\ket{1}\bra{1},
\\
&\sigma _{00|22}^\text{C}=\cos^2\theta\ket{0}\bra{0}, \sigma _{11|22}^\text{C}=\sin^2\theta\ket{1}\bra{1},
\end{split}
\end{equation}
where $\ket{\theta_{\pm}^{2}}=\cos\theta\ket{0}\pm\sin\theta\ket{1}$ and $\ket{\theta_{\pm}^{3}}=\cos\theta\ket{0}\pm i\sin\theta\ket{1}$. Components $\sigma _{01|22}^\text{C}$ and $\sigma _{10|22}^\text{C}$ do not exist as the probabilities of obtaining them are zero.

\subsection{Explicit form of assemblage $\Sigma_{ab|xy}^\text{dist}$}
The elements in assemblage after distillation are
\begin{equation}
\begin{split}
&\sigma_{00|00}^\text{dist}=\sigma_{11|00}^\text{dist}=\sigma_{01|11}^\text{dist}=\sigma_{10|11}^\text{dist}\\
&=P_\text{succ}\tilde{\sigma}_{00|00}+P_\text{fail}\sigma_{00|00}^\text{GGHZ}\\
&=\frac{1}{4}\begin{pmatrix}
\kappa^2\cos^2\theta & \kappa\cos\theta\sin\theta \\
\kappa\cos\theta\sin\theta &\sin^2\theta
\end{pmatrix}+\frac{(1-\kappa^2)\cos^2\theta}{4}\begin{pmatrix}
\cos^2\theta & \cos\theta\sin\theta \\
\cos\theta\sin\theta &\sin^2\theta
\end{pmatrix}\\
&=\frac{1}{4}\begin{pmatrix}
\kappa^2\cos^2\theta\sin^2\theta+\cos^4\theta  & \cos\theta\sin\theta(\kappa+\cos^2\theta-\kappa^2\cos^2\theta) \\
\cos\theta\sin\theta(\kappa+\cos^2\theta-\kappa^2\cos^2\theta) &\sin^2\theta(1+\cos^2\theta-\kappa^2\cos^2\theta)
\end{pmatrix}\\
\end{split}
\end{equation}
\begin{equation}
\begin{split}
&\sigma_{01|00}^\text{dist}=\sigma_{10|00}^\text{dist}=\sigma_{00|11}^\text{dist}=\sigma_{11|11}^\text{dist}\\
&=P_\text{succ}\tilde{\sigma}_{01|00}+P_\text{fail}\sigma_{01|00}^\text{GGHZ}\\
&=\frac{1}{4}\begin{pmatrix}
\kappa^2\cos^2\theta & -\kappa\cos\theta\sin\theta \\
-\kappa\cos\theta\sin\theta &\sin^2\theta
\end{pmatrix}+\frac{(1-\kappa^2)\cos^2\theta}{4}\begin{pmatrix}
\cos^2\theta & -\cos\theta\sin\theta \\
-\cos\theta\sin\theta &\sin^2\theta
\end{pmatrix}\\
&=\frac{1}{4}\begin{pmatrix}
\kappa^2\cos^2\theta\sin^2\theta+\cos^4\theta  & -\cos\theta\sin\theta(\kappa+\cos^2\theta-\kappa^2\cos^2\theta) \\
-\cos\theta\sin\theta(\kappa+\cos^2\theta-\kappa^2\cos^2\theta) &\sin^2\theta(1+\cos^2\theta-\kappa^2\cos^2\theta)
\end{pmatrix}\\
\end{split}
\end{equation}
\begin{equation}
\begin{split}
&\sigma_{00|01}^\text{dist}=\sigma_{11|01}^\text{dist}=\sigma_{00|10}^\text{dist}=\sigma_{11|10}^\text{dist}\\
&=P_\text{succ}\tilde{\sigma}_{00|01}+P_\text{fail}\sigma_{00|01}^\text{GGHZ}\\
&=\frac{1}{4}\begin{pmatrix}
\kappa^2\cos^2\theta & i\kappa\cos\theta\sin\theta \\
-i\kappa\cos\theta\sin\theta &\sin^2\theta
\end{pmatrix}+\frac{(1-\kappa^2)\cos^2\theta}{4}\begin{pmatrix}
\cos^2\theta & i\cos\theta\sin\theta \\
-i\cos\theta\sin\theta &\sin^2\theta
\end{pmatrix}\\
&=\frac{1}{4}\begin{pmatrix}
\kappa^2\cos^2\theta\sin^2\theta+\cos^4\theta  & i\cos\theta\sin\theta(\kappa+\cos^2\theta-\kappa^2\cos^2\theta) \\
-i\cos\theta\sin\theta(\kappa+\cos^2\theta-\kappa^2\cos^2\theta) &\sin^2\theta(1+\cos^2\theta-\kappa^2\cos^2\theta)
\end{pmatrix}\\
\end{split}
\end{equation}
\begin{equation}
\begin{split}
&\sigma_{01|01}^\text{dist}=\sigma_{10|01}^\text{dist}=\sigma_{01|10}^\text{dist}=\sigma_{10|10}^\text{dist}\\
&=P_\text{succ}\tilde{\sigma}_{01|01}+P_\text{fail}\sigma_{01|01}^\text{GGHZ}\\
&=\frac{1}{4}\begin{pmatrix}
\kappa^2\cos^2\theta & -i\kappa\cos\theta\sin\theta \\
i\kappa\cos\theta\sin\theta &\sin^2\theta
\end{pmatrix}+\frac{(1-\kappa^2)\cos^2\theta}{4}\begin{pmatrix}
\cos^2\theta & -i\cos\theta\sin\theta \\
i\cos\theta\sin\theta &\sin^2\theta
\end{pmatrix}\\
&=\frac{1}{4}\begin{pmatrix}
\kappa^2\cos^2\theta\sin^2\theta+\cos^4\theta  & -i\cos\theta\sin\theta(\kappa+\cos^2\theta-\kappa^2\cos^2\theta) \\
i\cos\theta\sin\theta(\kappa+\cos^2\theta-\kappa^2\cos^2\theta) &\sin^2\theta(1+\cos^2\theta-\kappa^2\cos^2\theta)
\end{pmatrix}\\
\end{split}
\end{equation}
\begin{equation}
\begin{split}
&\sigma _{00|02}^\text{dist}=\sigma _{10|02}^\text{dist}=\sigma_{00|12}^\text{dist}=\sigma_{10|12}^\text{dist}=\sigma_{00|20}^\text{dist}=\sigma _{01|20}^\text{dist}=\sigma _{00|21}^\text{dist}=\sigma _{01|21}^\text{dist}\\
&=P_\text{succ}\tilde{\sigma}_{00|02}+P_\text{fail}\sigma_{00|02}^\text{GGHZ}\\
&=\frac{1}{2}\begin{pmatrix}
\kappa^2\cos^2\theta & 0 \\
0 &0
\end{pmatrix}+\frac{(1-\kappa^2)\cos^2\theta}{2}\begin{pmatrix}
\cos^2\theta & 0 \\
0 &0
\end{pmatrix}=\frac{1}{2}\begin{pmatrix}
\kappa^2\cos^2\theta\sin^2\theta+\cos^4\theta & 0 \\
0 &0
\end{pmatrix}\\
\end{split}
\end{equation}
\begin{equation}
\begin{split}
&\sigma _{01|02}^\text{dist}=\sigma _{11|02}^\text{dist}=\sigma _{01|12}^\text{dist}=\sigma _{11|12}^\text{dist}=\sigma _{10|20}^\text{dist}=\sigma _{11|20}^\text{dist}=\sigma _{10|21}^\text{dist}=\sigma _{11|21}^\text{dist}\\
&=\frac{1}{2}\begin{pmatrix}
0  & 0 \\
0 &\sin^2\theta
\end{pmatrix}+\frac{(1-\kappa^2)\cos^2\theta}{2}\begin{pmatrix}
0 & 0 \\
0 &\sin^2\theta
\end{pmatrix}=\frac{1}{2}\begin{pmatrix}
0 & 0 \\
0 &\sin^2\theta(1+\cos^2\theta-\kappa^2\cos^2\theta)
\end{pmatrix}\\
\end{split}
\end{equation}
\begin{equation}
\begin{split}
&\sigma _{00|22}^\text{dist}=\begin{pmatrix}
\kappa^2\cos^2\theta\sin^2\theta+\cos^4\theta & 0 \\
0 &0
\end{pmatrix}, 
\sigma _{11|22}^\text{dist}=\begin{pmatrix}
0 & 0 \\
0 &\sin^2\theta(1+\cos^2\theta-\kappa^2\cos^2\theta)
\end{pmatrix}
\end{split}
\end{equation}
\subsection{Maximization of assemblage fidelity}
According to calculations
\begin{equation}
\begin{split}
\sum_{a, b\in\{0, 1\}} f(\sigma_{ab|xy}^\text{dist}, \sigma_{ab|xy}^\text{GHZ})=
\begin{cases}
\sqrt{\frac{1}{2}+\cos\theta\sin\theta(\cos^2\theta-\kappa^2\cos^2\theta+\kappa)} &x\neq2\,\ \text{and}\,\ y\neq2\\
\sqrt{\frac{\kappa^2\cos^2\theta\sin^2\theta+\cos^4\theta}{2}}+\sqrt{\frac{\sin^2\theta(1+\cos^2\theta-\kappa^2\cos^2\theta)}{2}} & x=2\,\ \text{or}\,\ y=2
\end{cases},
\end{split}
\end{equation}
it is easy to check 
\begin{equation}
\begin{split}
&\left[\sum_{a, b\in\{0, 1\}} f(\sigma_{ab|xy}^\text{dist}, \sigma_{ab|xy}^\text{GHZ})\right]^2-\left[\sum_{a, b\in\{0, 1\}} f(\sigma_{ab|22}^\text{dist}, \sigma_{ab|22}^\text{GHZ})\right]^2\\
=&\cos\theta\sin\theta\left[(1-\kappa^2)\cos^2\theta+\kappa-\sqrt{(1-\kappa^2)^2\cos^4\theta+(1+\kappa^2)(1-\kappa^2)\cos^2\theta+\kappa^2}\right]\\
\leq&\cos\theta\sin\theta\left[(1-\kappa^2)\cos^2\theta+\kappa-\sqrt{(1-\kappa^2)^2\cos^4\theta+2\kappa(1-\kappa^2)\cos^2\theta+\kappa^2}\right]\\
=&0.
\end{split}
\end{equation}
Then, we have the assemblage fidelity of 
\begin{equation}
\begin{split}
F_\text{2sDI}=\min_{x, y}\sum_{a, b\in\{0, 1\}} f(\sigma_{ab|xy}^\text{dist}, \sigma_{ab|xy}^\text{GHZ})&=\sqrt{\frac{1}{2}+\cos\theta\sin\theta(\cos^2\theta-\kappa^2\cos^2\theta+\kappa)}\\&\leq\sqrt{\frac{1}{2}+\cos\theta\sin\theta\left(\cos^2\theta+\frac{1}{4\cos^2\theta}\right)},
\end{split}
\end{equation}
with the equality holding when $\kappa=\frac{1}{2\cos^2\theta}$. 

\section{The derivation of $\kappa^\prime$}
The optimal filtering operation $C_\text{F}(\kappa^\prime)$ in regime of $N\to\infty$ in 1sDI scenario is the same as that in 2sDI scenario. We take the calculations in 2sDI scenario. Conditional upon a successful filtering, which occurs with probability
\begin{equation}
P_\text{succ}=\Tr\left[C_\text{F}(\kappa^\prime)\rho^{\text{C}} C_\text{F}^\dagger(\kappa^\prime)\right]=(\kappa^\prime)^2\cos^2\theta+\sin^2\theta,
\end{equation}
the assemblage $\Sigma_{ab|xy}^\text{GGHZ}$ is updated to $\tilde{\Sigma}_{ab|xy}=\{\tilde{\sigma}_{ab|xy}^\text{C}\}$ with 
\begin{equation}
\tilde{\sigma}_{ab|xy}^\text{C}=\frac{1}{P_\text{succ}}C_\text{F}(\kappa^\prime)\sigma^{\text{C}}_{ab|xy}C_\text{F}^\dagger(\kappa^\prime). 
\end{equation}
Here, $\rho^\text{C}=\Tr_\text{AB}(\ket{\text{GGHZ}_3}\bra{\text{GGHZ}_3})$. $\lim_{N\to\infty}P_\text{succ}^\text{2sDI}\to 1$ guarantees distillation of at least one copy of $\Sigma_{ab|xy}^\text{GHZ}$ in the asymptotic regime. The calculation of
\begin{equation}
\sum_{a, b}f\left(\tilde{\sigma}_{ab|xy}, \sigma_{ab|xy}^\text{GHZ}\right)=\sqrt{\frac{1}{2}\left(1+\frac{2\kappa^\prime\sin\theta\cos\theta}{(\kappa^\prime)^2\cos^2\theta+\sin^2\theta}\right)} \forall x, y
\end{equation}
lead to the assemblage fidelity
\begin{equation}\label{}
\begin{split}
F_\text{2sDI}(\tilde{\Sigma}_{ab|xy}, \Sigma_{ab|xy}^\text{GHZ})=\sqrt{\frac{1}{2}\left(1+\frac{2\kappa^\prime\sin\theta\cos\theta}{(\kappa^\prime)^2\cos^2\theta+\sin^2\theta}\right)}\leq 1. 
\end{split}
\end{equation}
The equality holds when $\kappa^\prime=\tan\theta$ so that the optimal filter operation in the regime of infinity copies $N$ is 
\begin{equation}\label{Eq:filter_infty}
C_\text{F}(\kappa^\prime)=\tan\theta\ket{0}\bra{0}+\ket{1}\bra{1}. 
\end{equation}
\end{document}